# Identifying optimal photovoltaic technologies for underwater applications

Jason A. Röhr[1,†], Ed Sartor[1], Joel N. Duenow[2], Zilun Qin[1,3], Juan Meng[1,3], Jason Lipton[1], Stephen A. Maclean[1], Udo Römer[4], Michael P. Nielsen[4], Suling Zhao[3], Jaemin Kong[5,6], Matthew O. Reese[2], Myles A. Steiner[2], N. J. Ekins-Daukes[4], *and* André D. Taylor[1,*,‡]

[1]Department of Chemical and Biomolecular Engineering, Tandon School of Engineering, New York University, Brooklyn, NY 11201, United States of America

[2]National Renewable Energy Laboratory, Golden, CO 80401, United States of America

[3]Key Laboratory of Luminescence and Optical Information, Beijing Jiaotong University, Ministry of Education, Beijing, 100044, China

[4]School of Photovoltaic and Renewable Energy Engineering, University of New South Wales, Sydney, New South Wales, Australia

[5]Research Institute for Green Energy Convergence Technology, Gyeongsang National University, Jinju, South Gyeongsang Province 52828, Republic of Korea

[6]Heeger Center for Advanced Materials and Research Institute for Solar and Sustainable Energies, Gwangju Institute of Science and Technology, Gwangju 61005, Republic of Korea

Improving solar energy collection in aquatic environments would allow for superior environmental monitoring and remote sensing, but the identification of optimal photovoltaic technologies for such applications is challenging as evaluation requires either field deployment or access to large water tanks. Here, we present a simple bench-top characterization technique that does not require direct access to water and therefore circumvents the need for field testing during initial trials of development. Employing LEDs to simulate underwater solar spectra at various depths, we compare Si and CdTe solar cells, two commercially available technologies, with GaInP cells, a technology with a wide band gap close to ideal for underwater solar harvesting. We use this method to show that while Si cells outperform both CdTe and GaInP cells under terrestrial AM1.5G solar irradiance, CdTe and GaInP cells outperform Si cells at depths > 2 m, with GaInP cells operating with underwater efficiencies approaching 54%.

†: jasonrohr@nyu.edu
‡: andre.taylor@nyu.edu

## Introduction

Long-term operation of devices for autonomous underwater vehicles (AUVs) and communication is limited by the lack of enduring and compact power sources and relies on power via a tether from a nearby source or from on-board batteries, increasing the architectural complexity or exceeding weight tolerances needed to maintain bouyancy.[1,2] Using solar cells to power underwater devices has been considered, but most attempts have had limited success due to the use of Si-based cells.[3,4] While highly efficient for terrestrial applications, Si has a narrow band gap of ~1.1 eV and relies on absorption of not only visible light, but also infrared (IR) light that suffers strong attenuation by water from scattering and absorption, resulting in a large decrease in power generation potential for underwater applications. So, while Si cells have been shown to produce useful power at shallow depths,[5–7] wider band-gap semiconductors should be employed for devices operating at greater depths, and identification or development of suitable technologies is necessary for this purpose.[8–11]

      Developing solar cells for underwater applications is complicated by the development of suitable materials and the complexity of characterizing such cells under aquatic light conditions. Generally, characterization of underwater solar cells requires large bodies of water, i.e., water tanks, seas, or lakes.[7,8] While characterizing solar cells in water tanks is relatively simple and can be done routinely,[5–7] the depth at which the cells can be measured is limited to the dimensions of the laboratory, typically a height of ~3 m. As measuring in deeper waters often requires traveling to locations where clear waters are found,[8] characterizing cells for deep-water applications is restrictive. Therefore, to

accelerate the development of underwater cells for such applications, convenient and accurate characterization methods that can be used within a typical laboratory setting must be developed.

Here, we present a simple and effective way of assessing underwater solar cells that does not require access to large bodies of water. Employing a light-emitting diode (LED) solar simulator, we mimic water-attenuated spectra by tuning the LED emission intensities. Using these simulated light conditions, we measured the power-conversion efficiencies ($\eta$) of crystalline (*c*)-Si, cadmium telluride (CdTe), and rear heterojunction gallium indium phosphide (*RHJ*-GaInP) solar cells. Our findings show that $\eta$ of the *c*-Si cell decreases from ~20.0% to 18.9% as the solar cell is illuminated by light simulating immersion into water from 0 to 9 m below sea level. Conversely, $\eta$ of the CdTe cell increases from 16.9% at 0 m to 24.1% at 2 m and subsequently decreases to 22.1% as the cell is immersed into deeper waters owing to its wider band gap of 1.4 eV. Finally, the *RHJ*-GaInP cell, with a terrestrial efficiency of 20-22%, can operate with $\eta > 50\%$ from 2 to 9 m below sea level due to its near-optimal band gap of 1.9 eV, which gives a better match to the spectral response of the cell when IR light is attenuated.

## Results & Discussion

### LED-based simulation of AM1.5G

Xenon (Xe) lamps are typically used to simulate sunlight due to the resemblance of its irradiance spectrum to that of the solar spectrum reaching Earth; however, solar simulators employing LEDs (here a VeraSol-2, class AAA; see Fig. S1 in the Supplementary Information at the end of this document) with peak wavelengths ranging from near ultraviolet (UV) to IR, can also be used to simulate solar irradiance spectra with high precision over the wavelength range relevant for underwater applications.[12-14] While the total power of a Xe lamp can be adjusted, the LED-based solar simulators offer more wavelength flexibility as the output power of each LED can be adjusted independently. Water-attenuated spectra can therefore be simulated using LEDs without the use of specially designed filters.

The AM1.5G solar spectrum, commonly used to characterize terrestrial solar cells, is shown in Fig. 1a, and the individual emission spectra of the LEDs we used to simulate this AM1.5G solar spectrum are shown in Fig. 1b. Each LED spectrum is labelled with the wavelength at peak intensity—ranging from 425 nm in the near UV to 1050 nm in the IR—all normalized to have an area of unity. The total spectral irradiance of the LED array, $\mathcal{E}_{\text{LED}}(\lambda)$, is given by the sum of the normalized LED spectra, $\varepsilon_i(\lambda)$, each weighted by a coefficient adjusting the magnitude of the *i*th LED peak, $a_i$,

$$\mathcal{E}_{\text{LED}}(\lambda) = \sum_{i=1} a_i \varepsilon_i(\lambda). \tag{1}$$

The incoming power density from a given light source, $p_{\text{in}}$, is then given as the integral of the spectral irradiance, $p_{\text{in}} = \int_{\lambda_1}^{\lambda_2} \mathcal{E}(\lambda)\, d\lambda$. Given the power density of a single LED as the integral of the irradiance of that LED, $p_{\text{in},i} = \int_{\lambda_1}^{\lambda_2} a_i \varepsilon_i(\lambda)\, d\lambda$, the total power density of the LED array is calculated as,

$$p_{\text{in,LED}} = \int_{\lambda_1}^{\lambda_2} \sum_{i=1} a_i \varepsilon_i(\lambda)\, d\lambda. \tag{2}$$

A comparison between the AM1.5G spectrum, the spectrum from a filtered Xe lamp, and the total LED emission spectrum, optimized to best represent the AM1.5G spectrum, are all shown in Fig. 1c. The $a_i$ coefficients used to obtain an optimum fit between the LEDs and the AMG1.5 spectrum are given in Table S1.

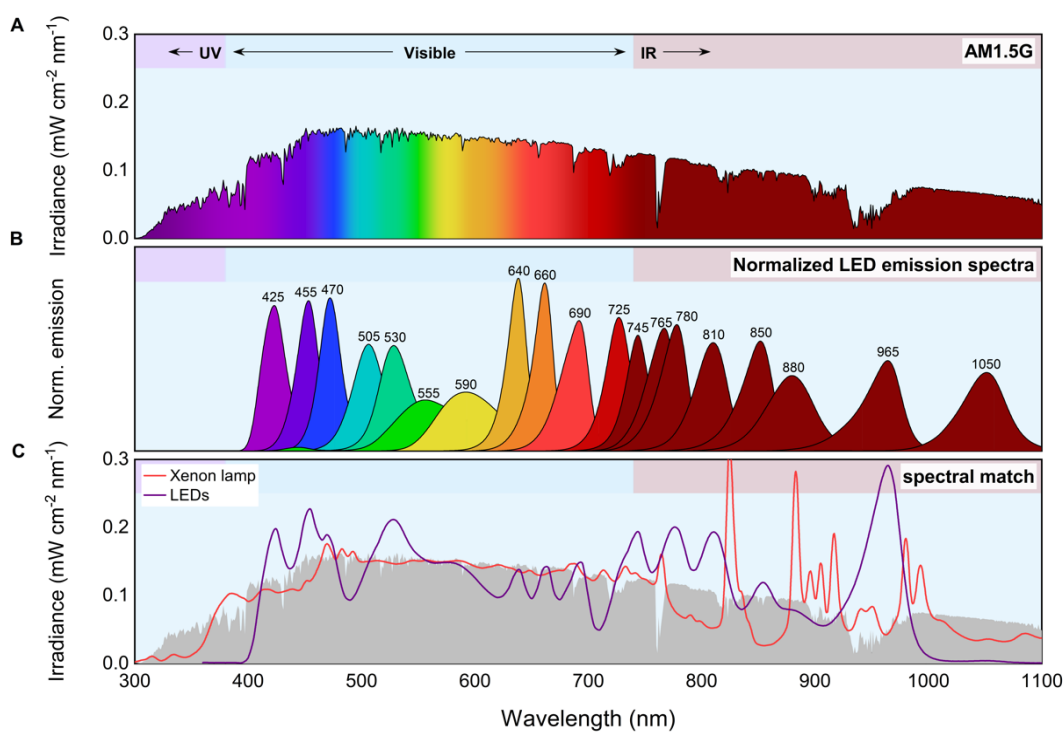

**Figure 1 – LED- and Xe-simulated sunlight. a**, AM1.5G solar spectrum from UV (300 nm) to IR (1100 nm). **b**, Emission spectra of the 19 LEDs used to simulate both AM1.5G and underwater light, ranging from 425 nm to 1050 nm. **c**, Comparison of irradiance spectra between AM1.5G, a Xe lamp with an AM1.5G filter and the combined LED spectrum by optimizing each LED to best match the AM1.5G solar spectrum.

Due to the absence of a UV LED in the array below 400 nm and a lower emission output of the IR LED at 1050 nm (Fig. 1b), parts of the LED spectrum show notable differences from that of the AM1.5G spectrum, while the Xe lamp simulates the entire AM1.5G spectrum with reasonable accuracy (Fig. 1c). Notwithstanding these differences, integrating all three irradiance spectra and comparing the resultant accumulative power density curves show that a good match is achieved across the compared wavelengths (Fig. 2a and Fig. S2a). Ultimately, the lack of a UV LED and the low output of the IR LED have little consequence when simulating underwater sunlight since water attenuates the solar irradiance in those wavelengths (Fig. S2b,c). Additionally, the spectral mismatches between the simulated sunlight and the reference spectra are accounted for mathematically, as explained below.

**Light attenuation in water**

As is evident from the water attenuation spectra shown in Fig. 2b, water scatters and absorbs light, especially in the UV and IR regions. While pure water absorbs the least across the visible spectrum (Fig. 2b),[15–17] sea and lake water can absorb a significant amount of visible light, depending on the level of salinity and concentration of organic matter (Fig. S3).[8,16–22] Waters off the coast of Key West were chosen for the present study as they represent relatively clear water, and their attenuation spectrum has been experimentally verified across the majority of visible region of interest. The Key West attenuation spectrum was not recorded between 300 nm to 360 nm and between 720 nm to 3000 nm; however, since a convergence in the spectra of Key West water and pure water was observed for $\lambda > 700$ nm, the pure water spectrum was used to extend the Key West spectrum to cover the entire 300 nm to 3000 nm range (dashed lines in Fig. 2b).[11]

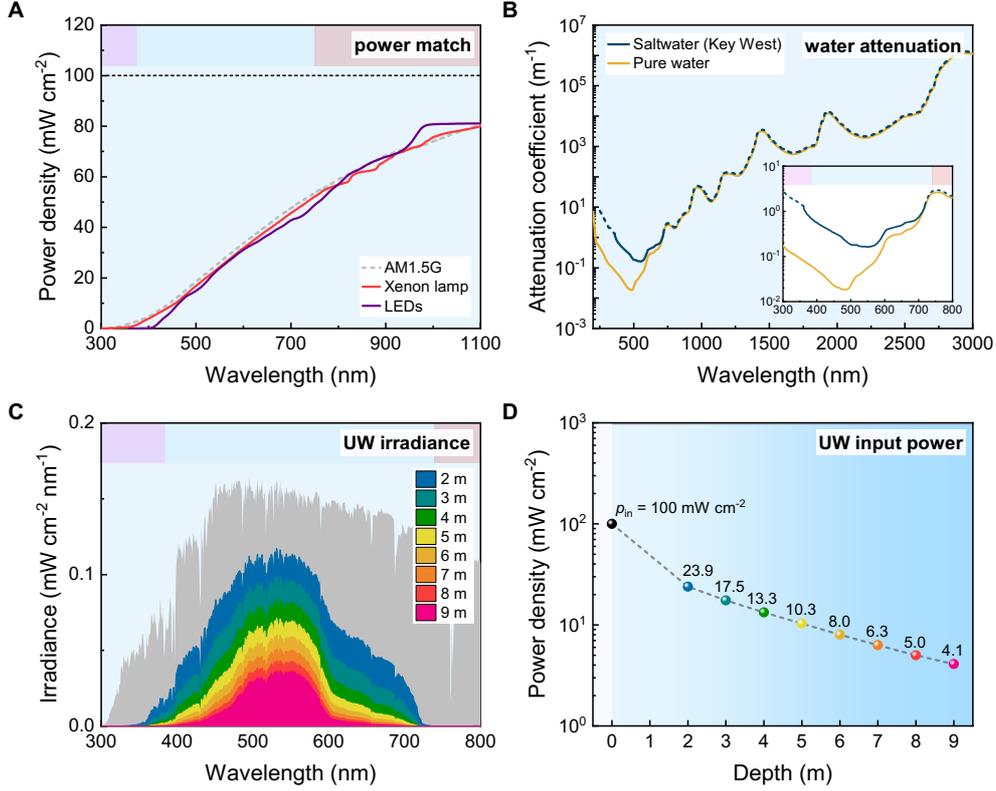

**Figure 2 – Water attenuation and resulting power. a**, Accumulated power density from integrating the irradiance spectra shown in (Fig. 1c). **b**, Attenuation coefficient spectra ($\alpha$) of salt water off the coast of Key West[8] and pure water.[15] Dashed lines indicate extensions made to the salt water spectrum using the pure water spectrum. Inset shows $\alpha$ values in a narrow range from 300 to 800 nm. **c**, Underwater irradiance spectra calculated using the Beer-Lambert law and the Key West attenuation spectrum. The AM1.5G spectrum is shown in grey for reference. **d**, Resultant input power density obtained from Eq. 3.

Underwater irradiance spectra, $\mathcal{E}_{\text{UW}}$, can be modelled by assuming a simple Beer-Lambert correction to the AM1.5G irradiance spectrum, $\mathcal{E}_{\text{UW}}(\lambda, D) = \mathcal{E}_{\text{AM1.5G}}(\lambda) e^{-\alpha(\lambda)D}$, where $\alpha(\lambda)$ is the wavelength-dependent attenuation spectrum of water (Fig. 2b) and $D$ is the water depth below sea level. As surface waves primarily affect the incoming solar spectrum a few meters below sea level and scattering of light tends to smooth out these fluctuations in deeper waters,[8] we investigated $D$ from 2 m and down while assuming that the water surface is relatively still. At those depths (> 2 m), all IR, and UV light is entirely absorbed, leaving predominately blue and green light. The calculated underwater irradiance spectra using the Key West $\alpha$ spectrum and $D$, ranging from 2 to 9 m, is shown in Fig. 2c.

In the same manner that the power density of the LED array was calculated, the available underwater power density, $p_{\text{in,UW}}$, can be calculated as,

$$p_{\text{in,UW}}(D) = \int_{\lambda_1}^{\lambda_2} \mathcal{E}_{\text{AM1.5G}}(\lambda) e^{-\alpha(\lambda)D} \, d\lambda. \tag{3}$$

For $D = 0$, the available power is simply that of the unperturbed AM1.5G spectrum, i.e., 100 mW cm$^{-2}$; however, as shown in Fig. 2d, the available power decreases rapidly from 23.9 mW cm$^{-2}$ to 4.1 mW cm$^{-2}$ as $D$ is varied from 2 to 9 m. Despite these low levels of transmitted power, we demonstrate that the underwater irradiance spectra can be simulated with high precision by tuning each LED in the solar simulator array (Fig. 1b).

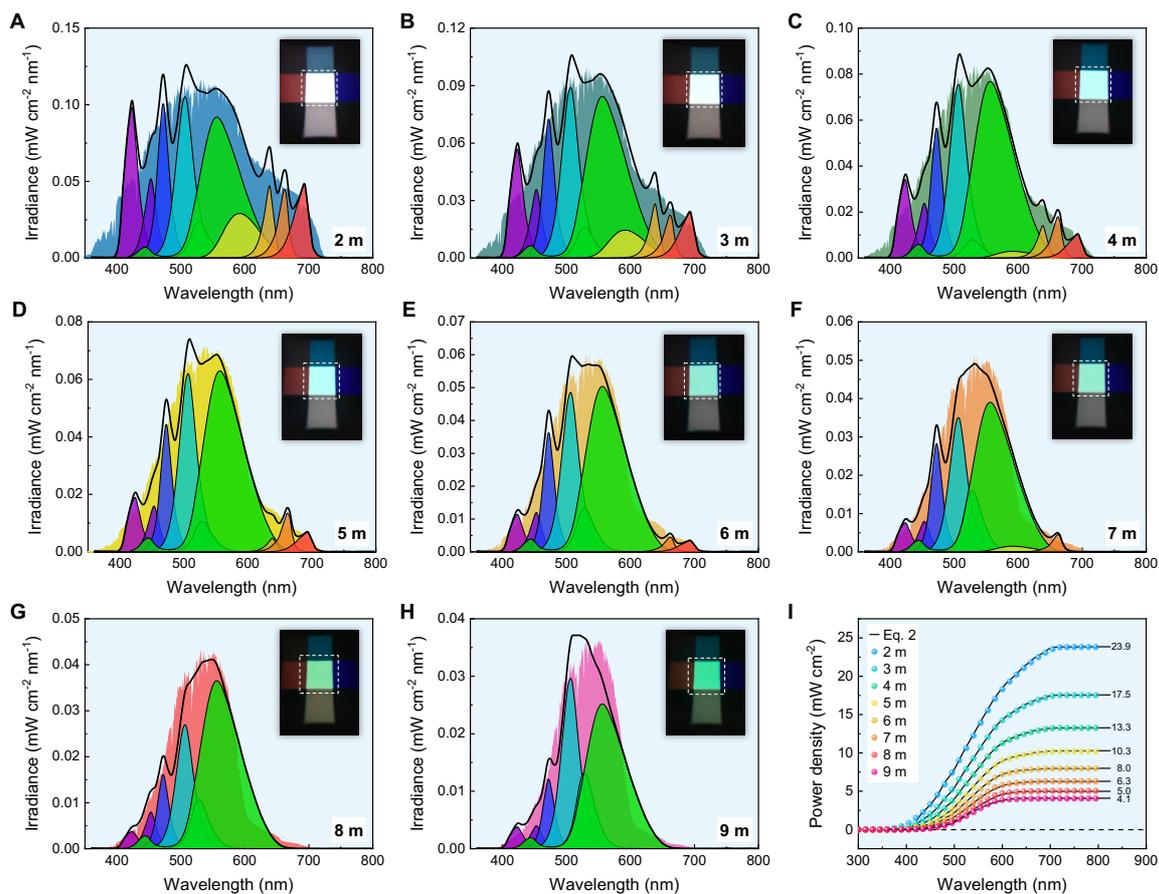

**Figure 3 – Simulations of underwater solar spectra. a-h,** Spectral match between individual LEDs (filled, colored LED emission curves) and the total LED-array emission (solid black line) compared with the Beer-Lambert-corrected AM1.5G spectra at depths from 2 to 9 m (colored spectra). The insets display photographs of the resulting LED light, showing the transition from white light at 0 m to blue-green light at 9 m. **i,** Accumulative power density curves obtained from integrating the total emission from the LEDs adjusted to best fit each underwater solar spectrum (solid black lines, Eq. 2) and the underwater-corrected AM1.5G solar spectra (filled circles, Eq. 3). The resulting power densities are listed.

## Simulation of underwater solar spectra

To simulate the underwater spectra shown in Fig. 2c, the power output from each LED in the array was varied to achieve a best fit to the underwater irradiance and the resulting accumulative power density curves, i.e., ensuring that the integrated irradiance spectra as given by Eq. 3 were as close to the LED power densities obtained from Eq. 2 for each depth. Each underwater spectrum, the spectral irradiance curves of each LED, and the total LED irradiance are shown in Fig. 3a-h for depths ranging between 2 and 9 m. Insets show photographs of the resulting LED light with the overall intensity decreasing with increasing depth (larger photographs are shown in Fig. S4). A shift from white to green light is visibly observed due to the attenuation of red light from the AM1.5G spectrum at simulated depths.

A comparison of the power density for the underwater spectra at each depth and their respective LED irradiance fits are shown in Fig. 3i. From this, it is evident that the LED array can simulate the underwater spectra with high precision. Since the intensity of light decreases with increased distance, $d$, from the simulator as $\sim d^{-2}$, the overall intensity of each measurement as a function of simulated device depth was determined by comparing the photocurrent of a Newport-calibrated Si reference cell to its calibrated value of 136.56 mA under the AM1.5G spectrum and accounting for the spectral mismatch between the AM1.5G and underwater spectra, as well as between the spectral responses of the test and reference cells.[23] Additional details are given in the Supplementary Information (Table S2).

**Identification and evaluation of underwater solar cell technologies**

With reliable characterization methods in place, we investigated the applicability of *c*-Si, CdTe, and *RHJ*-GaInP cells (Fig. S5-S8) by comparing their cell power outputs at their maximum power points, $p_{out} = \max(J \times V)$, along with their power conversion efficiencies,

$$\eta = \frac{p_{out}}{p_{in,LED}} \times 100\% = \frac{J_{SC}V_{OC}FF}{p_{in,LED}} \times 100\% \quad (4)$$

at simulated depths ranging from 0 to 9 m. In Eq. 4, $J_{SC}$ is the short-circuit current density, $V_{OC}$ is the open-circuit voltage and FF is the fill factor. We explicitly note that the efficiencies for $D > 0$ m are quoted relative to the power incident on the cell *at depth* ($p_{in,LED}$), rather than the AM1.5G spectrum at the water surface.[11] Solar cell fabrication details can be found in the Supplementary Information (Fig. S5-7).

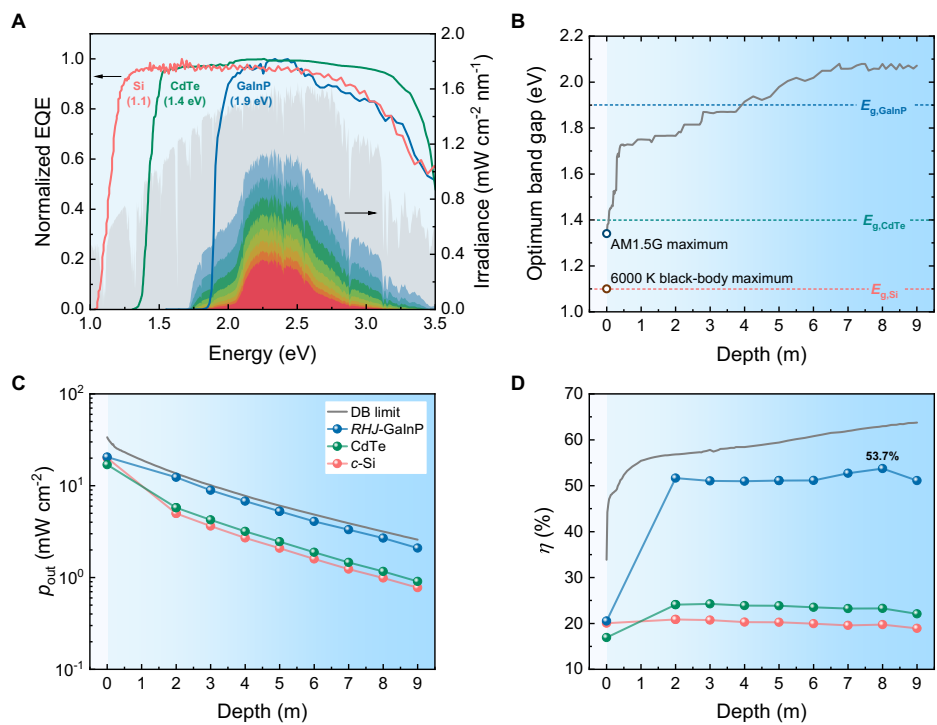

**Figure 4 – Identifying optimal technologies. a,** Normalized EQE spectra of *c*-Si, CdTe, and *RHJ*-GaInP cells along with the AM1.5G solar spectrum and the calculated underwater spectra. Band gap values are shown. **b,** Optimum band gap as a function of depth below sea level, calculated using a detailed-balance model, [11] showing the Si, CdTe, and GaInP solar cell band gaps for reference. The Shockley-Queisser limits assuming 6000 K black-body radiation and AM1.5G illumination at $D = 0$ m are shown. **c,** $p_{out}$ of the same cells under AM1.5G illumination ($D = 0$ m) and when simulating sunlight at $D = 2$-9 m. The grey solid line represents the maximum theoretical power density. **d,** Corresponding $\eta$ values, using values for $p_{in}$ given in Fig. 1g. The grey line represents the maximum theorical efficiency. Solar *J-V* curves are shown in Fig. S9a-c and tables with device characteristics are shown in Table *S3*.

Where Si and CdTe constitute commercially available technologies that future technologies for underwater solar cells must outcompete, *RHJ*-GaInP cells have wider band gaps and are often used as the top cell in high-efficiency multi-junction solar cells.[24–28] Due to their relatively small band gaps, both Si (1.1 eV) and CdTe (1.4 eV) solar cells are well-suited for absorbing across most of the AM1.5G solar spectrum, as evident by comparing external quantum efficiency (EQE) spectra with irradiance spectra (Fig. 4a); however, detailed-balance calculations (the methods used are discussed in a previous publication)[11] show that the optimum band gap for an ideal single-junction cell increases rapidly from 1.34 eV for AM1.5G illumination to ~1.72 eV at 0.4 m below sea level and continues to increase to ~2.1 eV for $D > 6$ m (Fig. 4b). As already proposed by Jenkins *et al.*,[8] as GaInP cells have larger band gaps (1.8-1.9 eV) and excellent overlap between the underwater irradiance spectra and the EQE spectrum (Fig. 4a), they should have the potential to outperform both Si and CdTe as similar $J_{SC}$ values are expected for all three technologies for $D \geq 2$ m while the $V_{OC}$ and FF of the GaInP cells are expected to greatly exceed that of the Si and CdTe cells, as is typical for solar cells with wider band gaps.

The measured $p_{out}$ from all three cells are shown in Fig. 4c, and their corresponding $\eta$ values are shown in Fig. 4d. *J-V* curves and characteristics are shown in Fig. S9a-c and Table S3, respectively. At sea level, under AM1.5G illumination, the *c*-Si cell generated $p_{out} = 20.0$ mW cm$^{-2}$ with an equivalent $\eta = 20.0\%$. When the *c*-Si cell was illuminated by simulated underwater light at depths varying from 2 to 9 m, $p_{out}$ rapidly decreased from 4.99 mW cm$^{-2}$ at 2 m to 0.78 mW cm$^{-2}$ at 9 m. Using the values for $p_{in,UW}$ as given by Eq. 3 (shown in Fig. 2d), $\eta$ for the *c*-Si cell increased to 20.9% at 2 m and subsequently decreased to 18.9% at 9 m, therefore retaining values for $\eta$ of 19-21% at the water surface and across all simulated depths. It should be noted that the c-Si cells measured here were chosen to be comparable with module-sized equivalents. So even though lab-scale Si solar cells with efficiencies as high as 26.1% have been demonstrated,[28] commercial silicon panels typically have efficiencies ~18-22%, similar to the Si cell measured here.

While generating less power under AM1.5G illumination (16.9 mW cm$^{-2}$) than the *c*-Si cell, the CdTe solar cell generated more power at simulated depths from 2 to 9 m (from 5.8 mW cm$^{-2}$ to 0.91 mW cm$^{-2}$). These higher power densities resulted in higher $\eta$ values ranging from 24.1% to 22.1% from 2 to 9 m, exceeding that of the *c*-Si cell at all simulated depths. CdTe cells could therefore be used as excellent alternatives to Si for underwater applications. It is here important to note that CdTe panels with $\eta = 19\%$ are commercially available (First Solar Series 6 CuRe).[29] These panels would likely yield higher underwater efficiencies than the CdTe cells investigated herein.

Under AM1.5G illumination, the $p_{out}$ and $\eta$ values of the *RHJ*-GaInP cell were similar to those of the *c*-Si cell, $p_{out} = 20.5$ mW cm$^{-2}$ and $\eta = 20.5\%$; however, under simulated underwater light, the *RHJ*-GaInP cell retained a very high power density ranging from 12.35 mW cm$^{-2}$ at 2 m to 2.10 mW cm$^{-2}$ at 9 m. These high power densities resulted in very high $\eta$ values in excess of 51% from 2 to 9 m, with a $\eta$ maximum approaching 54% at 8 m, which is close to the theoretical limit for a 1.9 eV band gap solar cell (Fig. 4d).[11] The high efficiencies of the *RHJ*-GaInP cell at depth is due to the very high $V_{OC}$ (from 1.45 V to 1.40 V between 2 and 9 m) and FF values (0.89 to 0.88), as a result of its 1.9 eV band gap, while generating similar $J_{SC}$ values to the *c*-Si and CdTe cells (from ~10 mA cm$^{-2}$ to ~1.7 mA cm$^{-2}$) as shown in Fig. S9. The improved operational depth and the remarkable efficiency improvement (more than 100% improvement) of the wide-band-gap GaInP cell over that of the two commercial technologies highlight the importance of using such semiconductor absorbers for underwater applications. It should be noted that while relatively large-area GaInP cells have been developed for aerospace applications, the manufacturing is specialized and volumes are low, resulting in high costs. Higher volume manufacturing and innovation for reducing the substrate cost is therefore required to reach costs comparable to CdTe.[30]

**Improvements to identified solar cell technologies and potential issues**

As estimated from suns-$V_{OC}$ measurements (Fig. S9d),[31] while the measured *c*-Si cell is operating with an ideality factor, $n$, close to unity, there is room for improvement for both the CdTe ($n = 1.53$) and *RHJ*-GaInP cells ($n = 1.12$). Record CdTe cell efficiencies have recently stalled at ~22% and presently suffer from non-radiative recombination that results in significant voltage losses;[28] however, the CdTe community has been working on refining the defect chemistry as well as improving band alignment at the front and back interfaces with aims of boosting efficiencies towards 28%.[32,33]

Even though the *RHJ*-GaInP cell greatly outperformed both the *c*-Si and CdTe cells at depth, it should be possible to increase the underwater efficiency of GaInP cells even more by increasing $E_g$ to more optimal values (2.0-2.1 eV), as has been done with lower-efficiency GaInP cells by alloying with Al.[25] CdTe technologies can similarly be modified to potentially improve underwater performance; Cd-Mg-Te alloys have been shown to have band gaps ranging from 1.5 to 3.5 eV depending on Mg content. While such Cd-Mg-Te alloys have mostly been leveraged in the fabrication of tandem cells,[34] or as back contact buffer layers for improved hole extraction in CdTe solar cells,[35] cells made using these as the main absorbers could be ideal candidates for underwater applications.

An important consideration is long-term operational stability of a solar panel while submerged in salt water. It has been shown that unencapsulated Si panels can remain submerged underwater during operation for months without any significant loss in $\eta$;[36] however, salt could be an issue for high-voltage installations. Moreover, cooling panels by submerging into water improves the efficiency. While not accounted for here, we have previously calculated that performance can be improved by up to ~10% if panels are submerged in sea water close to the freezing point.[11] While this thermal effect will be greater for *c*-Si due to its inferior temperature coefficient (-0.45 %/°C) relative to CdTe (-0.17 %/°C), spectral effects dominate such that CdTe modules should outperform Si underwater even when cooled. Finally, it is important to prohibit biofouling, i.e., the build-up of organic matter, on the solar cell encapsulant,[37] as this will result in power loss as light can no longer reach the active materials inside the solar cell.[38] Various antifouling approaches exist that can be explored for this purpose.[39]

## Conclusions

We demonstrated a facile and effective bench-top method for characterizing underwater solar cells that does not require submersion into water tanks. We reproduced solar irradiance spectra at varying depths using an LED solar simulator and used these to examine Si, CdTe, and GaInP solar cells. We found that GaInP solar cells outperform both Si and CdTe solar cells for underwater applications, with efficiencies approaching 54%. We subsequently discussed means for improving upon existing technologies to boost underwater performance. Previous methods for characterizing underwater solar cells relied on large water tanks for laboratory measurements or long-distance travel to take field measurements, thus our method reduces the effort necessary to characterize underwater solar cells, providing a facile path towards development of solar cells that can be used to power underwater technologies used for exploration and remote sensing.

## Acknowledgements

The authors are grateful for funding from New York University. We would also like to thank Dr. Miguel Modestino (NYU) for use of the LED solar simulator and Dr. David Young (NREL) for providing the *c*-Si cell. This work was co-authored by Alliance for Sustainable Energy, LLC, the manager and operator of the National Renewable Energy Laboratory for the U.S. Department of Energy (DOE) under Contract No. DE-AC36-08GO28308. The information, data, or work presented herein was funded, in part, by U.S. Department of Energy, Office of Energy Efficiency and Renewable Energy, Solar Energy Technology Office under agreements 34353 and 34358.

## Author Contributions

J.A.R. conceived the project; J.A.R., E.S., and M.A.S. performed the calculations; J.N.D. designed the CdTe devices; M.A.S. designed the GaInP devices; J.A.R., Z.Q. and J.M. characterized all devices with input from M.A.S.; J.L. and S.A.M. gathered all optical data; J.L., S.A.M., U.R., M.P.N., S.Z., J.K., and M.O.R. provided critical input during the discussions; J.A.R. wrote the initial manuscript with feedback from all other authors; J.A.R., M.O.R., M.A.S., N.J.E.-D., and A.D.T. oversaw the collaboration; A.D.T. supervised the project.

## Competing interests

The authors declare no competing interests

# Supplementary Information

The Supplementary Information contains a description of device fabrication processes, the procedures and equipment used to simulate sunlight, solar cell *J-V* curves and characteristics and mismatch correction calculations.

## S1. Solar cell characterization

<u>Device characterization:</u> All solar cell current density–voltage (*J-V*) curves were measured with a computer-controlled Keithley 2400 source measurement unit (*www.tek.com/keithley-source-measure-units/keithley-smu-2400-standard-series-sourcemeter*) under AM1.5G and different underwater solar spectra generated by a class AAA Verasol-2 LED-based solar simulator (*www.newport.com/p/VeraSol-2*). The light intensity of the simulator was calibrated using the Newport-certified reference multi-crystalline silicon cell (*www.newport.com/f/calibrated-reference-cell*). The individual LED intensities (Table S1) to generate underwater solar spectra were generated from literature-reported attenuation spectra (see Fig. S7 below) using a MatLab optimization script "SpectraMax.m" (*https://github.com/EdSartor/Solar*). All measurements were conducted in ambient air at room temperature.

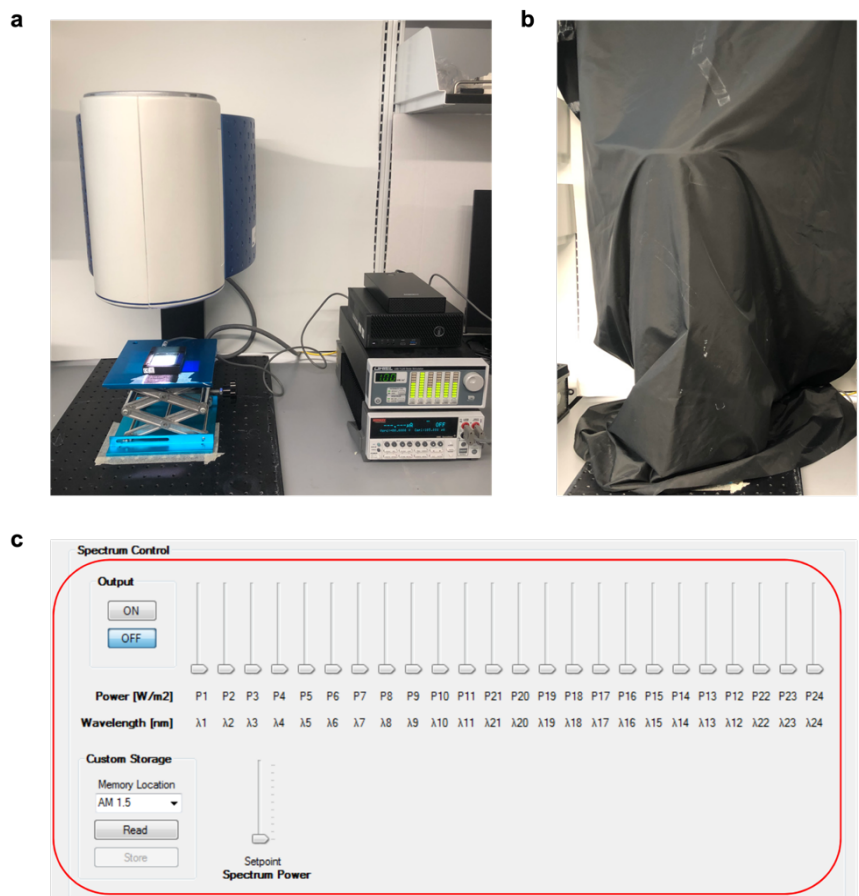

**Figure S1 – a,** The employed experimental setup consisting of the VeraSol-2 setup class AAA solar simulator coupled to a Keithley SMU. **b,** Same setup under light absorbing cloth to avoid ambient light interfering with the low-intensity measurements. **c,** Spectrum control software, allowing for fine tuning each individual LED in the setup.

| λ max (nm) | 425 | 455 | 470 | 505 | 530 | 555 | 590 | 640 | 660 | 690 | 725 | 745 | 765 | 780 | 810 | 850 | 880 | 965 | 1050 | |
|---|---|---|---|---|---|---|---|---|---|---|---|---|---|---|---|---|---|---|---|---|
| Depth (m) | λ max of integrated LED spectral irradiance peaks (mW cm$^{-2}$) | | | | | | | | | | | | | | | | | | | $p_\Sigma$ |
| 0 | 5.0 | 4.7 | 3.4 | 3.8 | 4.8 | 7.1 | 7.9 | 1.8 | 2.4 | 4.2 | 2.5 | 4.7 | 1.1 | 3.4 | 6.6 | 3.0 | 3.6 | 12.6 | 0.2 | **82.8** |
| 2 | 2.6 | 1.3 | 2.5 | 3.9 | 0.9 | 7.3 | 2.0 | 1.0 | 1.0 | 1.4 | 0 | 0 | 0 | 0 | 0 | 0 | 0 | 0 | 0 | **23.9** |
| 3 | 1.5 | 0.9 | 1.8 | 3.3 | 0.6 | 6.7 | 1.0 | 0.6 | 0.5 | 0.7 | 0 | 0 | 0 | 0 | 0 | 0 | 0 | 0 | 0 | **17.6** |
| 4 | 0.9 | 0.6 | 1.4 | 2.8 | 0.3 | 6.1 | 0.2 | 0.3 | 0.4 | 0.3 | 0 | 0 | 0 | 0 | 0 | 0 | 0 | 0 | 0 | **13.3** |
| 5 | 0.5 | 0.4 | 0.11 | 0.23 | 0.4 | 5.0 | 0 | 0.1 | 0.3 | 0.2 | 0 | 0 | 0 | 0 | 0 | 0 | 0 | 0 | 0 | **10.3** |
| 6 | 0.3 | 0.3 | 0.9 | 1.8 | 0.5 | 4.0 | 0 | 0 | 0.1 | 0.1 | 0 | 0 | 0 | 0 | 0 | 0 | 0 | 0 | 0 | **8.0** |
| 7 | 0.2 | 0.2 | 0.7 | 1.3 | 0.6 | 3.1 | 0.1 | 0 | 0.1 | 0 | 0 | 0 | 0 | 0 | 0 | 0 | 0 | 0 | 0 | **6.3** |
| 8 | 0.1 | 0.2 | 0.4 | 1.0 | 0.4 | 3.0 | 0 | 0 | 0 | 0 | 0 | 0 | 0 | 0 | 0 | 0 | 0 | 0 | 0 | **5.1** |
| 9 | 0.1 | 0.1 | 0.3 | 1.1 | 0.5 | 2.0 | 0 | 0 | 0 | 0 | 0 | 0 | 0 | 0 | 0 | 0 | 0 | 0 | 0 | **4.1** |

**Table S1** – $a$ coefficients (in dimensions of mW cm$^{-2}$) that determine the power density output of each LED in the array at a specific depth (0 to 9 m) along with the total power density of the array at a specific depth.

## S2. Power match and underwater irradiance spectra

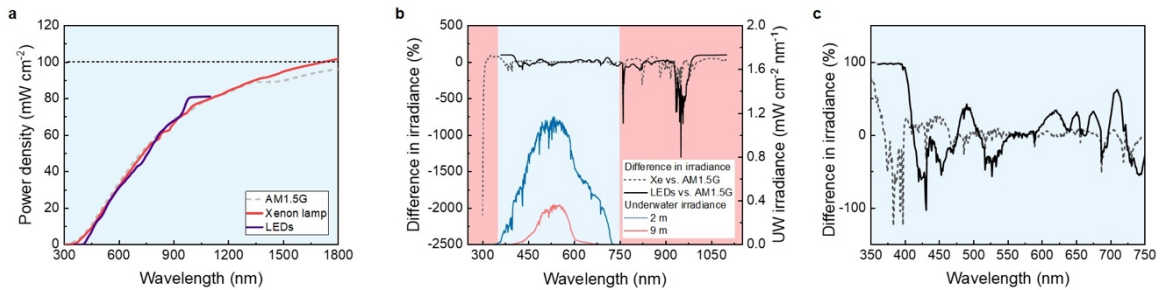

**Figure S2 – Power match and differences in irradiance spectra (%). a,** Accumulated power density from integrating the irradiance spectra shown in (Fig. 1c), showing how well the spectral match translates to a power match. **b,** Differences in irradiance (in %) between the AM1.5G and the LEDs and Xe lamp. The underwater irradiance spectra at 2 and 9 m are shown to highlight in which wavelength region the agreement between the LEDs and the AM1.5G is of highest importance. **c,** Difference in spectra shown from 350 to 750 nm.

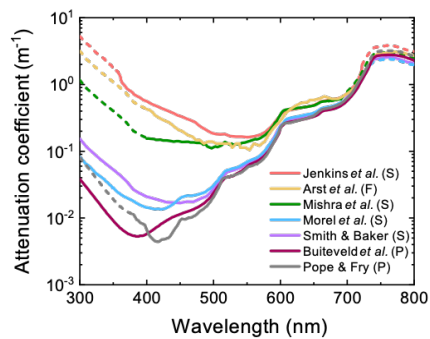

**Figure S3 - Attenuation spectra, either absorption spectra or diffuse attenuation coefficients of selected saltwater (S), fresh (F), and pure DI water (P).** Saltwater: Waters in the South Pacific (Morel et al.),[20] off the coast of Roatan Island in Honduras (Mishra et al.),[21] off the coast of Key West, Florida, in the United States (Jenkins et al.),[8] and waters from the Sargasso Sea (Smith and Baker),[19] Freshwater (lake water): From lake Äntu Sinijärv (Arst et al.).[22] Pure DI water: Buiteveld et al.[16] and Pope and Fry[17]. It should be noted that the spectrum by Pope and Fry was not recorded from 300 to 380 nm and the spectrum by Buiteveld et al. was used to extend the spectrum. Similarly, the spectra by Jenkins et al., Morel et al., Mishra et al., and Arst et al. were extended using the full spectrum by Smith and Baker. Extensions are shown as dashed lines. This data was previously published by Röhr et al.[11]

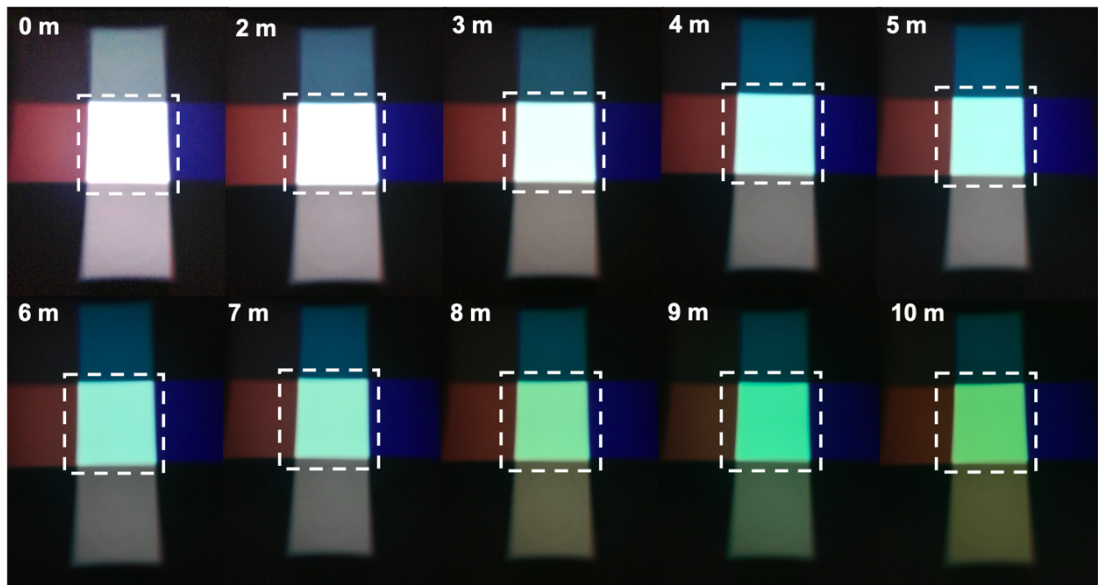

**Figure S4 – Photographs of resulting simulated underwater LED light.** Due to the absorption of IR and red light underwater, the colour of light shifts from white to green at large depths (here shown up to 10 m). This is simulated by varying each LED to best simulate sunlight penetrating water. These photographs are also shown as inset in Fig. 2. Photographs were taken with an Apple iPhone 8 without further editing besides brightness adjustments.

## S3. Solar cell fabrication

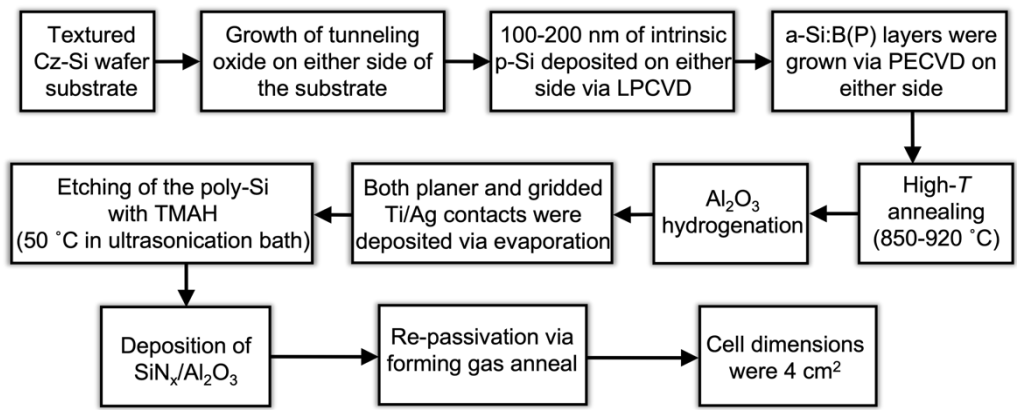

**Figure S5 –** Flowchart of the Si solar cell fabrication process.

<u>Crystalline ($c$)-Si solar cell:</u> The silicon cell was fabricated at NREL by first growing a tunnelling oxide on both sides of a double-side textured n-type Cz wafer at low temperature. 100–200 nm of intrinsic polycrystalline-Si was then deposited on both sides via low-pressure chemical-vapor deposition (LPCVD), and doped amorphous ($a$)-Si:B(P) layers were grown by plasma-enhanced chemical-vapor deposition (PECVD) on separate sides. The wafer was annealed at high temperature (850–920 °C) causing diffusion of dopants and crystallization of the $a$-Si layers. After an $Al_2O_3$ hydrogenation step, planer p-type Ti/Ag and gridded n-type Ti/Ag contacts were evaporated onto the sample, and the sample was dipped in TMAH (50 °C with ultrasonic agitation) to etch the poly-Si between the grid fingers. Finally, $SiN_x/Al_2O_3$ layers were deposited, and an FGA re-passivated the etched surface, forming a front/back poly-Si, rear junction device. The cell dimensions were 4 cm$^2$.

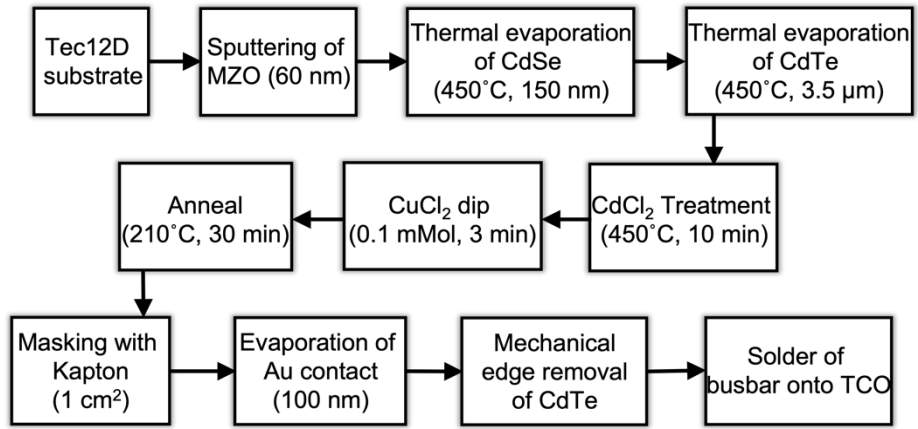

**Figure S6** – Flowchart of the CdTe solar cell fabrication process.

CdTe solar cells: The CdTe solar cell was fabricated at NREL on a 3" by 3" Pilkington Tec12D commercial substrates (3.2-mm-thick soda-lime glass/SnO$_2$:F/Undoped SnO$_2$). 60 nm MgZnO (4% MgO in ZnO) was deposited via RF magnetron sputtering at room temperature in 1% O$_2$/Ar ambient. 150 nm of CdSe was thermally evaporated while holding the substrate at 450 ºC, followed by a subsequent thermal evaporation of 3.5 $\mu$m of CdTe while holding the substrate at 450 ºC. The device was then exposed to CdCl$_2$ in a close-space sublimation chamber for 10 min, holding the substrate at 450 ºC in an H$_2$/N$_2$ ambient. The treated device was then copper doped by dipping in a 0.1 mM CuCl$_2$ solution for 3 minutes at room temperature, followed by annealing at 210 ºC for 30 min in lab air ambient. 1.5 cm by 1.5 cm devices were then cut from the larger coupon and a Kapton tape mask with 1 cm$^2$ exposed area was applied to the device. 100 nm of Au was deposited via thermal evaporation. The CdTe was mechanically etched from the transparent conducting oxide around the contact area, and an indium busbar was applied around the outside of the device. The front-side glass was apertured using a well-defined mask. Further details on the device fabrication method can be found in Ref 27.

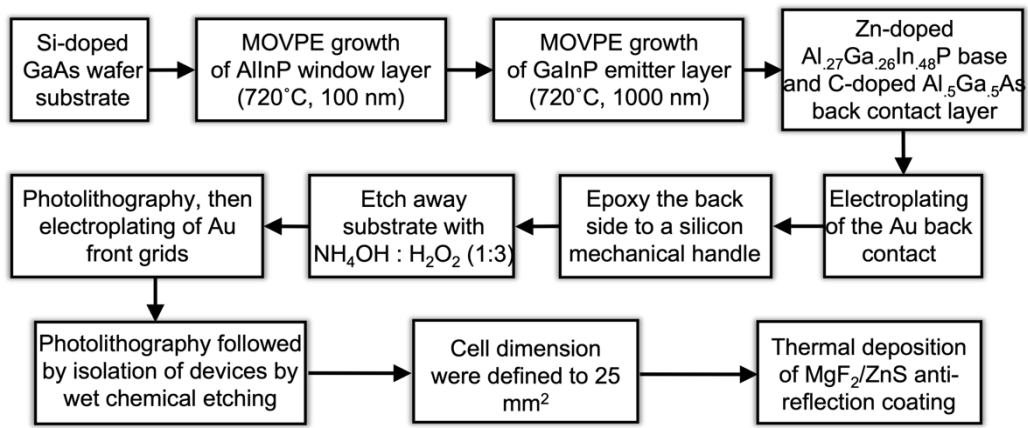

**Figure S7** – Flowchart of the GaInP solar cell fabrication process.

Rear-heterojunction (*RHJ*)-GaInP cells: The *RHJ*-GaInP cell was also grown at NREL using a custom-built MOVPE reactor. The cell was grown on a (100)-oriented silicon-doped GaAs wafer, miscut 2° toward <111>B. The cell was grown at 720 °C, with a nominal growth rate of ~3.9 $\mu$m/hr for the main absorber layer. The rear heterojunction architecture consists of an n-type AlInP window layer (~100 nm), a ~1 $\mu$m thick n-type (~$10^{17}$ cm$^{-3}$) silicon-doped GaInP emitter layer and a heterojunction with a higher bandgap p-type (~$2 \times 10^{17}$ cm$^{-3}$) Zn-doped Al$_{.27}$Ga$_{.26}$In$_{.48}$P base, and then a carbon-doped Al$_{.5}$Ga$_{.5}$As back contact layer. After the growth, individual cells were patterned with standard cleanroom processing techniques. A gold back contact was electroplated as the back contact layer, the semiconductor was secured to a handle and the substrate removed, gold grids were electroplated to the front and the cells were isolated with wet-chemical etchants. The cell dimensions were 25 mm$^2$. Further details can be found in Ref 24.

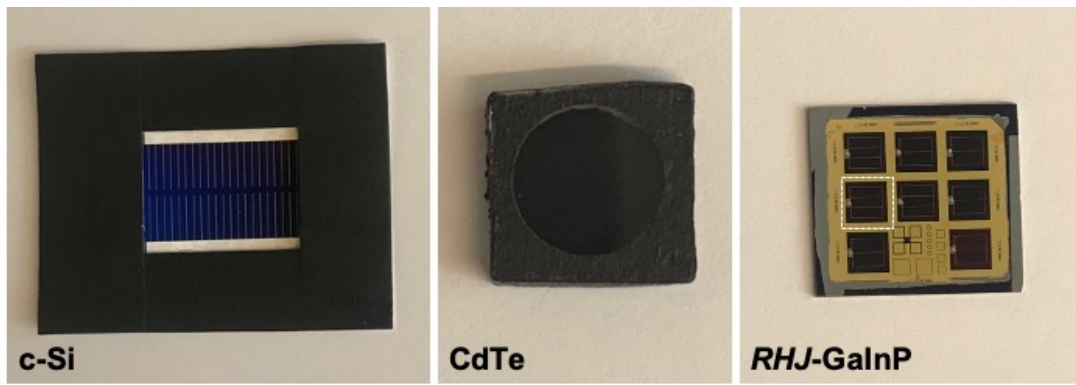

**Figure S8 – Photographs of studied solar cells.** Shaded *c*-Si and CdTe along with mesa isolated *RHJ*-GaInP solar cells. Photographs were taken with an Apple iPhone 8 without further editing.

## S4. Translation to reference conditions

A multi-crystalline silicon (*mc*-Si) cell was calibrated by Newport to have a short-circuit current of 136.56 mA under the AM1.5G reference spectrum (100 mW cm$^{-2}$), and this calibrated reference cell (RC) was used to set the total intensity of the LED simulator for each device that is tested by adjusting the height between the cell and the light source after setting the irradiance of the solar simulator to match the irradiance of either AM1.5G or the water-attenuated spectra. To use the RC to adjust the simulator intensity for measurement under an underwater spectrum one must account for i) the translation of the reference cell current to the new underwater reference spectrum, and ii) the spectral mismatch between the reference and test cells and between the underwater reference and simulator spectra. The methodology follows Ref. [23] and is summarized here, followed by some discussion points.

i) The quoted short-circuit current of 136.56 mA specifies the current of the *mc*-Si RC under the AM1.5G spectrum at 100 mW cm$^{-2}$. Next, we determine what the corresponding current would be under a particular underwater irradiance spectrum, i.e., at depths from 2 to 9 m. Denoting a depth-specific underwater irradiance spectrum as $\mathcal{E}_{UW}(\lambda, D)$ in units of mW cm$^{-2}$ nm$^{-1}$, where $D$ is the depth below sea level, the global reference spectrum as $\mathcal{E}_{AM1.5G}(\lambda)$, and the spectral response of the reference cell as $SR_{RC}(\lambda)$, the expected current of the reference cell under the underwater spectrum is given by Eq. 18 in Ref. 23,

$$I_{SC,UW}^{RC}(D) = 136.56 \text{ mA} \times \frac{\int \mathcal{E}_{UW}(\lambda,D)d\lambda}{100} \frac{\int \mathcal{E}_{AM1.5G}(\lambda)SR_{RC}(\lambda)d\lambda}{\int \mathcal{E}_{UW}(\lambda,D)SR_{RC}(\lambda)d\lambda} \frac{100.33}{\int \mathcal{E}_{UW}(\lambda,D)d\lambda} \quad \text{S1}$$

Note that the spectral response of the solar cell is related to the cell's corresponding external quantum efficiency (EQE) as,

$$SR_{RC}(\lambda) = \frac{\lambda}{hc} EQE(\lambda). \quad \text{S2}$$

The integrals are over the entire wavelength range. For the global spectrum, the reference cell is calibrated at 100 mW cm$^{-2}$ whereas the total irradiance in the spectrum is 100.33 mW cm$^{-2}$. For the underwater spectrum, $\int \mathcal{E}_{UW}(\lambda, D)d\lambda$ represents both the measurement irradiance and the total irradiance. Table S2 gives the calculated values of the current of the mc-Si solar cell under the calculated underwater spectra.

**Table S2** - Spectral correction calculations for the three test cells (*c*-Si, CdTe and *RHJ*-GaInP). The value of 136.56 mA (denoted with an * in the first row) is the calibration value as reported by Newport, for the AM1.5G spectrum at 100 mW m$^{-2}$ and the numbers listed in that row are the currents the silicon should measure under the LED spectra given the cells spectral response. Mismatch-correction values for the three test cells are shown along with the corrected $I_{SC}$ values for the mc-Si cell used for the solar simulator calibration.

| Underwater | | *mc*-Si (cert.) | *c*-Si | | CdTe | | *RHJ*-GaInP | |
|---|---|---|---|---|---|---|---|---|
| Depth (m) | Power (mW cm$^{-2}$) | $I_{SC,UW}$ (mA) | $M$ (-) | $I_{SC,corrected}$ (mA) | $M$ (-) | $I_{SC,corrected}$ (mA) | $M$ (-) | $I_{SC,corrected}$ (mA) |
| 0 | 100.00 | 136.56* | 0.990 | 137.95 | 0.975 | 140.10 | 0.870 | 157.01 |
| 2 | 23.81 | 34.63 | 1.001 | 34.38 | 0.998 | 34.50 | 1.006 | 34.24 |
| 3 | 17.54 | 25.28 | 1.001 | 25.25 | 0.998 | 25.33 | 1.000 | 25.27 |
| 4 | 13.25 | 19.04 | 1.001 | 19.03 | 0.998 | 19.08 | 1.000 | 19.04 |
| 5 | 10.20 | 14.63 | 1.000 | 14.63 | 0.997 | 14.68 | 0.993 | 14.73 |
| 6 | 7.97 | 11.43 | 1.000 | 11.42 | 0.998 | 11.45 | 0.995 | 11.48 |
| 7 | 6.30 | 9.04 | 1.000 | 9.04 | 0.997 | 9.06 | 1.000 | 9.04 |
| 8 | 5.03 | 7.22 | 1.000 | 7.22 | 0.998 | 7.23 | 0.988 | 7.31 |
| 9 | 4.05 | 5.81 | 1.000 | 5.81 | 0.998 | 5.82 | 0.986 | 5.90 |

ii) The second calculation is to determine the spectral mismatch correction factor, *M*, for each test device under each underwater spectrum. This allows for the adjustment of the reference cell current before the measurement. Denoting the underwater LED simulator spectrum as $\mathcal{E}_{LED}(\lambda, D)$ and the spectral response of the test cell (device under test, DUT) as $SR_{DUT}(\lambda)$, following the same calculation is describes by S2, the correction factor is given by Eq. 14 in Ref. 23,

$$M(D) = \frac{\int \mathcal{E}_{LED}(\lambda,D) SR_{DUT}(\lambda) d\lambda \times \int \mathcal{E}_{UW}(\lambda,D) SR_{RC}(\lambda) d\lambda}{\int \mathcal{E}_{UW}(\lambda,D) SR_{DUT}(\lambda) d\lambda \times \int \mathcal{E}_{LED}(\lambda,D) SR_{RC}(\lambda) d\lambda}. \qquad S3$$

Table S2 gives the values of *M* for each cell and spectrum. Finally, the RC $I_{SC}$ must therefore be adjusted via Eq. 15 in Ref. [23] to:

$$I^{RC}_{SC,measured} = \frac{I^{RC}_{SC,UW}}{M} \qquad S4$$

where $I^{RC}_{SC,measured}$ is the measured reference *mc*-Si reference cell current under the simulator and $I^{RC}_{SC,UW}$ is the current of the the *mc*-Si reference cell as calcuated in equation S1. The values that the *mc*-Si cell must be set to are shown in Table S2. The corrected current density of 15.735 mA cm$^{-2}$ for the *RHJ*-GaInP cell at 100 mW cm$^{-2}$ matches the certified measurement current density of that cell to within ~5%.[40]

We note that this methodology is mathematically precise but in practice must be applied with some caution. We assume that the the photocurrent varies linearly with intensity, meaning that if the intensity doubles then so does the $I_{SC}$. This is a good assumption near one-sun intensities and higher, but not always reliable at very low intensities where trap states can photofill. The overall correction factor is as much as ~1/25 for the measurements under the 9 m underwater spectrum. That would likely be too large a correction for a certified efficiency measurement, and instead a different reference cell would be chosen. But given the difficulty of doing a primary calibration underwater and the unavailabilty of higher bandgap calibrated reference cells, we deem the corrected estimates of the efficiencies as reasonable. Finally, we note that the values of *M* for the three test cells are all very close to each other, which indicates a good simulation of the underwater spectrum. The overall correction is dominated by the recalibration of the reference cell from the global spectrum to the underwater spectra.

## S5. Solar cell *J-V* curves and characteristics

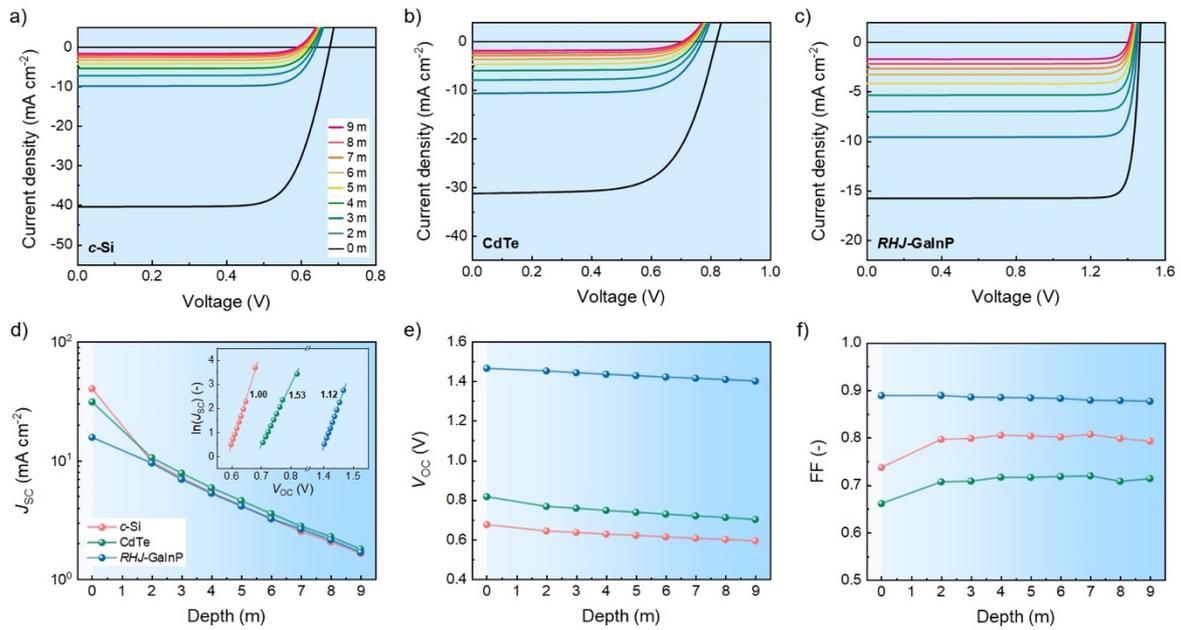

**Figure S9 – Solar cell *J-V* curves and device characteristics. a,** *c*-Si cell. **b,** CdTe cell. **c,** *RHJ*-GaInP cell. **d,** $J_{SC}$ as a function of device depth. Inset shows suns-$V_{OC}$ measurements and extraction of ideality factor ($n$) using Eq. S5. **e,** $V_{OC}$ as function of depth. **f,** FF as a function of depth.

| | *c*-Si | | | | | CdTe | | | | | *RHJ*-GaInP | | | | |
|---|---|---|---|---|---|---|---|---|---|---|---|---|---|---|---|
| Depth (m) | $J_{SC}$ (mA cm$^{-2}$) | $V_{OC}$ (V) | FF (-) | $P_{out}$ (mW cm$^{-2}$) | $\eta$ (%) | Depth (m) | $J_{SC}$ (mA cm$^{-2}$) | $V_{OC}$ (V) | FF (-) | $P_{out}$ (mW cm$^{-2}$) | $\eta$ (%) | Depth (m) | $J_{SC}$ (mA cm$^{-2}$) | $V_{OC}$ (V) | FF (-) | $P_{out}$ (mW cm$^{-2}$) | $\eta$ (%) |
| 0 | 40.317 | 0.678 | 0.738 | 20.184 | 20.184 | 0 | 31.205 | 0.819 | 0.662 | 16.922 | 16.922 | 0 | 15.735 | 1.466 | 0.889 | 20.511 | 20.511 |
| 2 | 9.840 | 0.646 | 0.797 | 5.070 | 21.212 | 2 | 10.613 | 0.770 | 0.708 | 5.785 | 24.202 | 2 | 9.556 | 1.453 | 0.890 | 12.351 | 51.680 |
| 3 | 7.204 | 0.639 | 0.799 | 3.676 | 20.886 | 3 | 7.877 | 0.761 | 0.709 | 4.251 | 24.154 | 3 | 6.985 | 1.444 | 0.886 | 8.936 | 50.774 |
| 4 | 5.401 | 0.630 | 0.806 | 2.745 | 20.636 | 4 | 5.928 | 0.750 | 0.717 | 3.189 | 23.976 | 4 | 5.332 | 1.436 | 0.886 | 6.780 | 50.977 |
| 5 | 4.215 | 0.624 | 0.804 | 2.115 | 20.533 | 5 | 4.632 | 0.740 | 0.717 | 2.459 | 23.877 | 5 | 4.165 | 1.429 | 0.885 | 5.266 | 51.125 |
| 6 | 3.261 | 0.617 | 0.802 | 1.613 | 20.165 | 6 | 3.601 | 0.731 | 0.719 | 1.893 | 23.657 | 6 | 3.256 | 1.422 | 0.884 | 4.091 | 51.139 |
| 7 | 2.549 | 0.609 | 0.808 | 1.255 | 19.914 | 7 | 2.822 | 0.722 | 0.720 | 1.467 | 23.293 | 7 | 2.667 | 1.416 | 0.880 | 3.322 | 52.722 |
| 8 | 2.083 | 0.603 | 0.799 | 1.004 | 19.686 | 8 | 2.302 | 0.714 | 0.709 | 1.165 | 22.842 | 8 | 2.170 | 1.409 | 0.879 | 2.687 | 52.696 |
| 9 | 1.663 | 0.597 | 0.794 | 0.787 | 19.199 | 9 | 1.810 | 0.704 | 0.714 | 0.910 | 22.206 | 9 | 1.704 | 1.402 | 0.877 | 2.096 | 51.110 |

**Table S3 –** Solar cell characteristics of the *c*-Si, CdTe and *RHJ*-GaInP cells under AM1.5G illimination and under simulated underwater light from 2 to 9 m (assuming water off the coast of Key West).

The ideality factors shown in the inset of Fig. S5d were calculated using the following equation by fitting to ln($J_{SC}$) as a function of $V_{OC}$.[31]

$$n = \frac{q}{k_B T}\left(\frac{d \ln J_{SC}}{d V_{OC}}\right)^{-1} \qquad \text{S5}$$

# References

1. Wang, X. *et al*. Reviews of Power Systems and Environmental Energy Conversion for Unmanned Underwater Vehicles. *Renew. Sustain. Energy Rev.* **16**, 1958–1970 (2012).
2. Delphin Raj, K. M. *et al*. Underwater network management system in internet of underwater things: Open challenges, benefits, and feasible solution. *Electron.* **9**, 1–33 (2020).
3. Arima, M., Okashima, T. & Yamada, T. Development of a Solar-Powered Underwater Glider. *2011 IEEE Symp. Underw. Technol. UT'11 Work. Sci. Use Submar. Cables Relat. Technol. SSC'11* 1–5 (2011). doi:10.1109/UT.2011.5774120
4. Joshi, K. B., Costello, J. H. & Priya, S. Estimation of Solar Energy Harvested for Autonomous Jellyfish Vehicles (AJVs). *IEEE J. Ocean. Eng.* **36**, 539–551 (2011).
5. Lanzafame, R. *et al*. Field Experience with Performances Evaluation of a Single-Crystalline Photovoltaic Panel in an Underwater Environment. *IEEE Trans. Ind. Electron.* **57**, 2492–2498 (2010).
6. Sheeba, K. N., Rao, R. M. & Jaisankar, S. A Study on the Underwater Performance of a Solar Photovoltaic Panel. *Energy Sources, Part A Recover. Util. Environ. Eff.* **37**, 1505–1512 (2015).
7. Enaganti, P. K., Dwivedi, P. K., Sudha, R., Srivastava, A. K. & Goel, S. Underwater Characterization and Monitoring of Amorphous and Monocrystalline Solar Cells in Diverse Water Settings. *IEEE Sens. J.* **20**, 2730–2737 (2020).
8. Jenkins, P. P. *et al*. High-Bandgap Solar Cells for Underwater Photovoltaic Applications. *IEEE J. Photovoltaics* **4**, 202–206 (2014).
9. Walters, R. J. *et al*. Multijunction Organic Photovoltaic Cells for Underwater Solar Power. *2015 IEEE 42nd Photovolt. Spec. Conf. PVSC 2015* 1–3 (2015). doi:10.1109/PVSC.2015.7355644
10. Kong, J. *et al*. Underwater Organic Solar Cells via Selective Removal of Electron Acceptors near the Top Electrode. *ACS Energy Lett.* **4**, 1034–1041 (2019).
11. Röhr, J. A., Lipton, J., Kong, J., Stephen, A. & Taylor, A. D. Efficiency Limits of Underwater Solar Cells. *Joule* **4**, 1–10 (2020).
12. Bliss, M., Wendlandt, S., Betts, T. R. & Gottschalg, R. Towards a High Power, all LED Solar Simulator Closely Matching Realistic Solar Spectra. *24th Eur. Photovolt. Sol. Energy Conf.* 3321–3326 (2009).
13. Kolberg, D., Schubert, F., Lontke, N., Zwigart, A. & Spinner, D. M. Development of Tunable Close Match LED Solar Simulator with Extended Spectral Range to UV and IR. *Energy Procedia* **8**, 100–105 (2011).
14. Linden, K. J., Neal, W. R. & Serreze, H. B. Adjustable Spectrum LED Solar Simulator. *Proc. SPIE 9003, Light. Diodes Mater. Devices, Appl. Solid State Light. XVIII* **9003**, 900317 (2014).
15. Segelstein, D. J. The Complex Refractive Index of Water. (1981).
16. Buiteveld, H., Hakvoort, J. H. M. & Donze, M. Optical Properties of Pure Water. *Ocean Opt. XII* **2258**, 174–183 (1994).
17. Pope, R. M. & Fry, E. S. Absorption Spectrum (300-700 nm) of Pure Water. II. Integrating Cavity Measurements. *Appl. Opt.* **36**, 8710-8723 ST-Absorption spectrum (380-700 nm) (1997).
18. Wozniak, B. & Dera, J. *Light Absorption in Sea Water. Springer* **33**, (2007).
19. Smith, R. C. & Baker, K. S. Optical properties of the natural waters (200-800 nm). *Appl. Opt.* **20**, 177–184 (1981).
20. Morel, A. *et al*. Optical properties of the 'clearest' natural waters. *Limnol. Oceanogr.* **52**, 217–229 (2007).

21. Mishra, D. R., Narumalani, S., Rundquist, D. & Lawson, M. Characterizing the Vertical Diffuse Attenuation Coefficient for Downwelling Irradiance in Coastal Waters: Implications for Water Penetration by High Resolution Satellite Data. *ISPRS J. Photogramm. Remote Sens*. **60**, 48–64 (2005).
22. Arst, H. *et al*. Optical Properties of Boreal Lake Waters in Finland and Estonia. *Boreal Environ. Res*. **13**, 133–158 (2008).
23. Osterwald, C. R. Translation of device performance measurements to reference conditions. *Sol. Cells* **18**, 269–279 (1986).
24. Geisz, J. F., Steiner, M. A., García, I., Kurtz, S. R. & Friedman, D. J. Enhanced external radiative efficiency for 20.8% efficient single-junction GaInP solar cells. *Appl. Phys. Lett*. **103**, (2013).
25. Perl, E. E. *et al*. Development of high-bandgap AlGaInP solar cells grown by organometallic vapor-phase epitaxy. *IEEE J. Photovoltaics* **6**, 770–776 (2016).
26. Geisz, J. F. *et al*. Six-junction III–V solar cells with 47.1% conversion efficiency under 143 Suns concentration. *Nat. Energy* **5**, 326–335 (2020).
27. Ablekim, T. *et al*. Thin-Film Solar Cells with 19% Efficiency by Thermal Evaporation of CdSe and CdTe. *ACS Energy Lett*. **5**, 892–896 (2020).
28. Green, M. A. *et al*. Solar Cell Efficiency Tables (Version 58). *Prog. Photovoltaics Res. Appl*. **29**, 657–667 (2021).
29. First Solar Series 6 CuRe. Available at: https://www.firstsolar.com/en/Modules/Series-6-CuRe.
30. Horowitz, K. A., Remo, T. W., Smith, B. & Ptak, A. J. A Techno-Economic Analysis and Cost Reduction Roadmap for III-V Solar Cells. *Tech. Rep*. NREL/TP-6A20-72103 (2018).
31. Gunawan, O., Gokmen, T. & Mitzi, D. B. Suns- VOC characteristics of high performance kesterite solar cells. *J. Appl. Phys*. **116**, (2014).
32. Kanevce, A., Reese, M. O., Barnes, T. M., Jensen, S. A. & Metzger, W. K. The roles of carrier concentration and interface, bulk, and grain-boundary recombination for 25% efficient CdTe solar cells. *J. Appl. Phys*. **121**, 214506 (2017).
33. Duenow, J. N. & Metzger, W. K. Back-surface recombination, electron reflectors, and paths to 28% efficiency for thin-film photovoltaics: A CdTe case study. *J. Appl. Phys*. **125**, 053101 (2019).
34. Mathew, X. *et al*. Development of a semitransparent CdMgTe/CdS top cell for applications in tandem solar cells. *Semicond. Sci. Technol*. **24**, 015012 (2009).
35. Bothwell, A. M., Drayton, J. A. & Sites, J. R. Efficiency Advances in Thin CdSeTe/CdTe Solar Cells with CdMgTe at the Back. *Conf. Rec. IEEE Photovolt. Spec. Conf*. **2020**-**June**, 1248–1253 (2020).
36. Rosa-Clot, M., Rosa-Clot, P., Tina, G. M. & Scandura, P. F. Submerged photovoltaic solar panel: SP2. *Renew. Energy* **35**, 1862–1865 (2010).
37. Crimmins, D. M. *et al*. Long-endurance test results of the solar-powered AUV system. in *OCEANS 2006 IEEE* 1–5 (2006).
38. Ageev, M. D., Blidberg, D. R., Jalbert, J., Melchin, C. J. & Troop, D. P. Results of the evaluation and testing of the solar powered AUV and its subsystems. in *Proceedings of the IEEE Symposium on Autonomous Underwater Vehicle Technology* 137–145 (2002). doi:10.1109/auv.2002.1177216
39. Cao, S., Wang, J. D., Chen, H. S. & Chen, D. R. Progress of marine biofouling and antifouling technologies. *Chinese Sci. Bull*. **56**, 598–612 (2011).
40. Green, M. *et al*. Solar cell efficiency tables (version 57). *Prog. Photovoltaics Res. Appl*. **29**, 3–15 (2021).